\documentclass[aps,pra,preprint,amsmath,amssymb]{revtex4-2}
\usepackage[utf8]{inputenc}
\usepackage{graphicx}
\usepackage{hyperref}
\hypersetup{colorlinks=true,linkcolor = blue, citecolor = blue, urlcolor = blue}
\usepackage{dcolumn}
\usepackage{color}
\usepackage{braket}
\usepackage{textcomp} 
\usepackage{natbib} 
\begin{document}		

\title{Non-monotonic dependence of OAM Schmidt spectrum on crystal thickness}  

\author{Harshal Jain}
\email{jainharshal52@gmail.com}
\affiliation{Department of Physics,
Indian Institute of Technology Kanpur, Kanpur,
UP 208016, India}

\author{Suman Karan}
\affiliation{Department of Physics,
Indian Institute of Technology Kanpur, Kanpur,
UP 208016, India}

\author{Radhika Prasad}
\affiliation{Department of Physics,
Indian Institute of Technology Kanpur, Kanpur,
UP 208016, India}
\author{Anand K. Jha}
\email{akjha9@gmail.com}
\affiliation{Department of Physics,
Indian Institute of Technology Kanpur, Kanpur,
UP 208016, India}

\pagestyle{plain}
\begin{abstract}
The orbital angular momentum (OAM) of photons provides a high-dimensional resource for quantum information protocols. The dimensionality of OAM-entangled states generated via spontaneous parametric down-conversion (SPDC) is quantified by the angular Schmidt spectrum.  Here, we experimentally investigate the dependence of the angular Schmidt spectrum on the thickness of the nonlinear crystal.  Contrary to previous studies reporting a monotonic decrease in the Schmidt number with increasing crystal thickness, we report the first experimental observation of a non-monotonic behavior, as we demonstrate an increase in the Schmidt number beyond a certain crystal thickness. We attribute this to the spatial walk-off effect in the anisotropic nonlinear crystal and explain it using a theoretical model that is devoid of standard phase-matching approximations. These findings can have important implications for high-dimensional entangled state generation.

\end{abstract}
\maketitle
\section{Introduction}
It is well established that a photon in a Laguerre–Gaussian mode $LG_p^l(\rho,\phi)$ carries orbital angular momentum (OAM) quantized in integer multiples of $\hbar$, where $l$ and $p$ denote the azimuthal and radial mode indices, respectively~\cite{allen1992orbital}. The OAM of photons offers a discrete and infinite-dimensional Hilbert space~\cite{barnett1990quantum,yao2006fourier}. Consequently, OAM-entangled states generated via spontaneous parametric down-conversion (SPDC) serve as an important resource for high-dimensional quantum information protocols~\cite{wang2017generation,hu2018beating,jha2011supersensitive,luo2019quantum}. Such states offer several advantages, including enhanced information capacity and improved robustness against noise~\cite{cerf2002security,ecker2019overcoming,zhu2021high}. For a Gaussian pump beam, the dimensionality of the OAM entanglement is quantified through the angular Schmidt spectrum~\cite{law2004analysis,torres2003quantum,di2010measurement}, defined as the probability of detecting signal and idler photons with OAM $l\hbar$ and $-l\hbar$, respectively. The corresponding Schmidt number quantifies the effective dimensionality of the OAM-entangled state~\cite{nielsen2000quantum}. Therefore, accurate determination of the Schmidt spectrum is essential for reliable characterization of high-dimensional OAM entanglement. 

Several theoretical approaches have been developed to characterize the angular Schmidt spectrum of OAM-entangled states. Torres \textit{et al.} derived one of the earliest formulations for calculating the spectrum in collinear geometry~\cite{torres2003quantum}. Their treatment employed an approximate phase-matching function that neglected the contribution of spatial walk-off. Within this framework, the angular Schmidt number was found to decrease monotonically with increasing crystal thickness. Later, Miatto \textit{et al.} derived expressions for the coincidence amplitudes in both thick- and thin-crystal regimes using a similar approximation for the phase-matching function~\cite{miatto2011full}. The angular Schmidt spectrum can be obtained from these expressions by summing over the radial modes of the signal and idler fields; however, the resulting formulation involves infinite summations over radial modes and suffers from convergence issues. A detailed discussion of these formulations and their limitations is provided in Appendix~\ref{appendix:approximation}. Similar coincidence-amplitude expressions were also derived by Yao \textit{et al.} under the thin-crystal and collinear approximations, with the phase-matching function taken to be unity~\cite{yao2011angular}. While the angular Schmidt spectrum was not considered in that work, its calculation from the coincidence amplitudes likewise requires infinite radial-mode summations and faces similar convergence challenges. The approximate coincidence-amplitude formulations discussed above have been widely adopted in subsequent studies of high-dimensional OAM entanglement. They have been used to investigate maximally entangled states, entanglement concentration, radial-mode correlations, hyperentanglement, remote state preparation, and quantum state characterization~\cite{liu2018coherent,xu2022manipulating,zhang2013high,lu2015orbital,salakhutdinov2012full,chen2012comblike,wu2016electro,su2014remote,zhang2018violation,xu2024efficient}. In addition, these formulations have been widely cited as the available theoretical description of the two-photon OAM spectrum~\cite{romero2012increasing,krenn2013entangled,roger2013non,krenn2014generation,zhang2014simulating,plick2015physical,bolduc2016direct,ritboon2017proposed,erhard2018twisted,luo2019chip,cao2020distribution,kysela2020path,baghdasaryan2021justifying,schwaller2022optimizing,karan2023postselection,sevilla2024spectral}. Their widespread use and recognition make it important to assess the validity of the underlying phase-matching approximation.

Subsequently, Miatto \textit{et al.} employed a geometric argument to study the angular Schmidt spectrum and derived an approximate $1/\sqrt{L}$ scaling of the Schmidt number with crystal thickness~\cite{miatto2012bounds}. This result implies a monotonic reduction in the effective dimensionality of the OAM-entangled state with increasing crystal thickness. A similar behavior was later obtained using a Gaussian approximation to the phase-matching function~\cite{miatto2012cartesian}. More recently, Kulkarni \textit{et al.} derived a formulation for the angular Schmidt spectrum that retains the complete phase-matching function and accounts for all radial modes~\cite{kulkarni2018angular}. The resulting expression is applicable to both collinear and non-collinear geometries and avoids explicit summation over radial modes. However, their analysis was restricted to thin crystals, for which the Schmidt number also exhibited a monotonic decrease with increasing crystal thickness. Thus, all previous theoretical studies reported a monotonic decrease of the Schmidt number with increasing crystal thickness~\cite{torres2003quantum,miatto2012bounds,miatto2012cartesian,kulkarni2018angular}, while its behavior beyond the thin-crystal regime using the complete phase-matching function remains largely unexplored. Additionally, the dependence of the angular Schmidt spectrum on crystal thickness has not been investigated experimentally.

In this paper, we report experimental observations of non-monotonic dependence of the Schmidt number on crystal thickness, where the Schmidt number initially decreases and subsequently increases. 

\section{Theory}
\subsection{Angular Schmidt spectrum and Schmidt number}
To investigate the dependence of the angular Schmidt spectrum on crystal thickness, we employ the formulation derived in Ref.~\cite{kulkarni2018angular}, which retains the complete phase-matching function and accounts for all radial-mode contributions. 
After tracing over the radial degrees of freedom, the OAM-entangled two-photon state can be written in the Schmidt-decomposed form~\cite{law2004analysis,torres2003quantum,jha2011partial}
\begin{equation}\label{eq}
|\psi_2\rangle = \sum_{l}\sqrt{S_l}\left|l\right\rangle_s\left|-l\right\rangle_i.
\end{equation} 

The angular Schmidt spectrum \(S_l\) represents the probability of detecting signal \((s)\) and idler \((i)\) photons in the OAM eigenstates \(|l\rangle_s\) and \(|-l\rangle_i\), corresponding to OAM values \(l\hbar\) and \(-l\hbar\), respectively. The angular Schmidt spectrum is given by~\cite{torres2003quantum,miatto2011full,yao2011angular,jha2011partial,kulkarni2018angular}
\begin{align}\label{eq:schmidt_spectrum}
S_l &\propto \iint_0^\infty
\Bigg|
\iint_{-\pi}^{\pi}
V(\rho_s,\rho_i,\phi_s,\phi_i)
\nonumber\\
&\qquad\times
\Phi(\rho_s,\rho_i,\phi_s,\phi_i,L,\theta_p)
e^{il(\phi_s-\phi_i)}
\, d\phi_s \, d\phi_i
\Bigg|^2
\nonumber\\
&\qquad\times \rho_s\rho_i\,d\rho_s\,d\rho_i.
\end{align}
Here,
$(\rho_s\cos\phi_s,\rho_s\sin\phi_s)$ and
$(\rho_i\cos\phi_i,\rho_i\sin\phi_i)$ denote the transverse momenta of the signal and idler photons, respectively. The parameters $L$ and $\theta_p$ denote the crystal thickness and phase-matching angle. The function $V(\rho_s,\rho_i,\phi_s,\phi_i)$ describes the transverse momentum distribution of the pump beam, while $\Phi(\rho_s,\rho_i,\phi_s,\phi_i,L,\theta_p)$ represents the phase-matching function. Since this formulation retains the complete phase-matching function and fully accounts for radial-mode contributions, it is applicable to both collinear and non-collinear geometries and does not rely on the thin-crystal approximation.

The effective dimensionality of the OAM-entangled state is quantified by the Schmidt number,
\begin{equation}
\label{eq:schmidt_number}
K = \frac{1}{\sum_l S_l^2},
\end{equation}
where $S_l$ is normalized such that \(\sum_l S_l = 1\). The crystal-thickness dependence of the angular Schmidt spectrum is governed by the phase-matching function. In the following subsection, we examine the phase-matching function and show how its exact and approximate forms lead to qualitatively different predictions for the Schmidt spectrum.

\subsection{Phase-matching function and the effect of spatial walk-off on the angular Schmidt spectrum}
As evident from Eq.~(\ref{eq:schmidt_spectrum}), the crystal-thickness dependence of the angular Schmidt spectrum is governed by the phase-matching function. We therefore examine the phase-matching function and the longitudinal phase mismatch between the pump, signal, and idler fields, which together with the crystal thickness determine its form inside the nonlinear crystal. The phase-matching function is given by~\cite{hong1985theory,walborn2010spatial,karan2020phase}
\begin{align}
\Phi(\mathbf{q}_s,\mathbf{q}_i,L,\theta_p)
&=
\int_{-L}^{0}
\exp(-i\Delta k_z z)\,dz
\nonumber\\
&=
\mathrm{sinc}\left(\frac{L\Delta k_z}{2}\right)
e^{-iL\Delta k_z/2},
\label{eq:phasematching}
\end{align}
where
\begin{equation}
\Delta k_z = k_{sz}+k_{iz}-k_{pz}.
\label{eq:expression_delta_kz}
\end{equation}

The phase-matching function therefore depends on both the crystal thickness \(L\) and the longitudinal phase mismatch \(\Delta k_z\). The quantities \(k_{jz}\) denote the longitudinal components of the wave vectors, where \(j=p,s,i\) correspond to the pump, signal, and idler fields, respectively. The corresponding transverse wave vectors are denoted by
\(\mathbf q_p=(q_{px},q_{py})\),
\(\mathbf q_s=(q_{sx},q_{sy})\),
and
\(\mathbf q_i=(q_{ix},q_{iy})\). The dependence on the phase-matching angle \(\theta_p\) enters through these longitudinal wave-vector components inside the nonlinear crystal.
For type-I phase matching in a uniaxial birefringent crystal, the longitudinal wave-vector components are given by~\cite{rubin1996transverse,walborn2010spatial,karan2020phase}
\begin{align}\label{eq:k's}
k_{pz} &= -\alpha_p q_{px} + \eta_p \frac{\omega_{p0}}{c}
- \frac{c\left(\beta_p^2 q_{px}^2 + \gamma_p^2 q_{py}^2\right)}{2\eta_p \omega_{p0}}, \nonumber\\
k_{sz} &= n_{so} \frac{\omega_{s0}}{c}
- \frac{c}{2 n_{so} \omega_{s0}} \left(q_{sx}^2 + q_{sy}^2\right), \nonumber\\
k_{iz} &= n_{io} \frac{\omega_{i0}}{c}
- \frac{c}{2 n_{io} \omega_{i0}} \left(q_{ix}^2 + q_{iy}^2\right).
\end{align}

The coefficients associated with the pump beam are given by
\begin{align}\label{pump-coeff}
\alpha_p &= \frac{(n_{po}^2 - n_{pe}^2)\sin\theta_p \cos\theta_p}
{n_{po}^2 \sin^2\theta_p + n_{pe}^2 \cos^2\theta_p}, \nonumber\\
\beta_p &= \frac{n_{po} n_{pe}}
{n_{po}^2 \sin^2\theta_p + n_{pe}^2 \cos^2\theta_p}, \nonumber\\
\gamma_p &= \frac{n_{po}}
{\sqrt{n_{po}^2 \sin^2\theta_p + n_{pe}^2 \cos^2\theta_p}}, \nonumber\\
\eta_p &= \frac{n_{po} n_{pe}}
{\sqrt{n_{po}^2 \sin^2\theta_p + n_{pe}^2 \cos^2\theta_p}}.
\end{align}

Here, $n_{jo}$ and $n_{je}$ denote the ordinary and extraordinary refractive indices, respectively. The central frequencies of the pump, signal, and idler fields are denoted by $\omega_{p0}$, $\omega_{s0}$, and $\omega_{i0}$.

To investigate the role of spatial walk-off in determining the crystal-thickness dependence of the angular Schmidt spectrum, we compare the exact phase mismatch given by Eqs.~(\ref{eq:expression_delta_kz}) and (\ref{eq:k's}) with the approximate phase mismatch employed in previous studies~\cite{torres2003quantum,miatto2011full,yao2011angular,miatto2012bounds,miatto2012cartesian}. For comparison with the approximate phase mismatch, it is convenient to express the exact phase mismatch as
\begin{equation}
\Delta k_z^{\mathrm{exact}}
=
\Delta k_z^{\mathrm{approx}}
+
\Delta k_z^{\mathrm{extra}},
\label{eq:deltak_decomposition}
\end{equation}
where
\begin{subequations}
\label{eq:deltak_terms}
\begin{align}
\Delta k_z^{\mathrm{approx}}
&=
\frac{|\mathbf q_s-\mathbf q_i|^2}
{2|\mathbf k_p|},
\label{eq:deltak_approx}
\\
\Delta k_z^{\mathrm{extra}}
&=
\frac{n_{so}\omega_{s0}}{c}
+
\frac{n_{io}\omega_{i0}}{c}
-
\frac{\eta_p\omega_{p0}}{c}
+
\alpha_p q_{px}.
\label{eq:deltak_walkoff}
\end{align}
\end{subequations}

Since $\beta_p$ and $\gamma_p$ remain close to unity over a wide range of phase-matching-angle, we set $\beta_p=\gamma_p=1$ to facilitate comparison with the approximate phase mismatch. 
The quantity $\Delta k_z^{\mathrm{approx}}$ is the conventional phase-mismatch expression obtained under the assumptions of collinear phase matching and identical refractive indices for the pump, signal, and idler fields inside the nonlinear crystal~\cite{torres2003quantum,miatto2011full,yao2011angular,miatto2012bounds,miatto2012cartesian}. 
The remaining contribution, $\Delta k_z^{\mathrm{extra}}$, contains both the residual phase-mismatch term and the linear term $\alpha_p q_{px}$ associated with the birefringent propagation of the extraordinary pump beam, which accounts for spatial walk-off inside the crystal.

Consequently, the principal difference between the exact and approximate phase-matching formulations is captured by the additional contribution $\Delta k_z^{\mathrm{extra}}$, which is neglected in the approximate treatment.
\begin{figure}[h!]
    \centering
    \includegraphics[
    width=\columnwidth,
    trim=7 8 6 7,
    clip
    ]{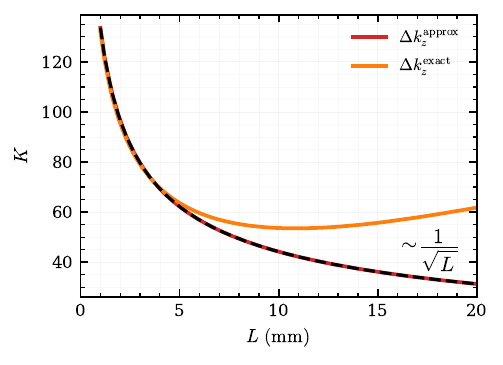}
    \caption{\textbf{Dependence of the Schmidt number on crystal thickness.}
    The red curve corresponds to the approximate phase mismatch, for which the Schmidt number decreases approximately as \(1/\sqrt{L}\)~\cite{miatto2012bounds}. The orange curve corresponds to the exact phase mismatch evaluated at \(\theta_p = 32.910^\circ\), exhibiting a non-monotonic dependence on crystal thickness due to spatial walk-off. The black dashed line indicates the \(1/\sqrt{L}\) scaling.}
    \label{fig:kvsL_spectrum_fitted_curve}
\end{figure}
Having identified the additional contribution to the exact phase mismatch, we now investigate its effect on the angular Schmidt spectrum by numerically evaluating the Schmidt number using both the exact and approximate phase-matching conditions. Figure~\ref{fig:kvsL_spectrum_fitted_curve} shows the dependence of the Schmidt number on crystal thickness for both phase-matching models. When the approximate phase mismatch is used, the Schmidt number decreases monotonically with crystal thickness and follows an approximate $(1/\sqrt{L})$ scaling, consistent with the prediction of Ref.~\cite{miatto2012bounds}. In contrast, when the exact phase mismatch is used, the Schmidt number exhibits a non-monotonic dependence on crystal thickness. Initially, the Schmidt number decreases with increasing crystal thickness, similar to the approximate case. However, beyond a certain crystal thickness, the Schmidt number begins to increase. 

Although \(\Delta k_z^{\mathrm{extra}}\) contains both a residual constant term and the spatial walk-off contribution, the qualitative change in the Schmidt-number dependence is primarily caused by the spatial walk-off term, whose influence becomes increasingly significant as the crystal thickness increases.
Consequently, the exact phase-matching model predicts a behavior that is qualitatively different from that obtained using the approximate phase mismatch. These results demonstrate that spatial walk-off plays a crucial role in determining the crystal-thickness dependence of the angular Schmidt spectrum. 

The limitations of the approximate phase-matching model extend beyond the crystal-thickness dependence of the Schmidt number. As discussed in Appendix~\ref{appendix:approximation}, although the approximate phase-matching model reproduces the previously reported agreement between the thick- and thin-crystal formulations~\cite{miatto2011full}, it fails to predict the zero-radial-mode probability distribution obtained using the complete phase-matching function (Appendix~\ref{appendix:zero_mode}). Furthermore, the angular Schmidt spectrum calculated using the approximate formulation does not converge with increasing radial-mode truncation and becomes progressively flatter, demonstrating that the approximate phase-matching model leads to qualitatively incorrect predictions for the angular Schmidt spectrum (Appendix~\ref{appendix:schmidt_comparison}). These results further emphasize the importance of retaining the complete phase-matching function for accurately describing high-dimensional OAM entanglement.
\begin{figure*}[t!]
    \centering
    \includegraphics[
    width=\textwidth,
    trim= 6 8 2 5,
    clip
    ]{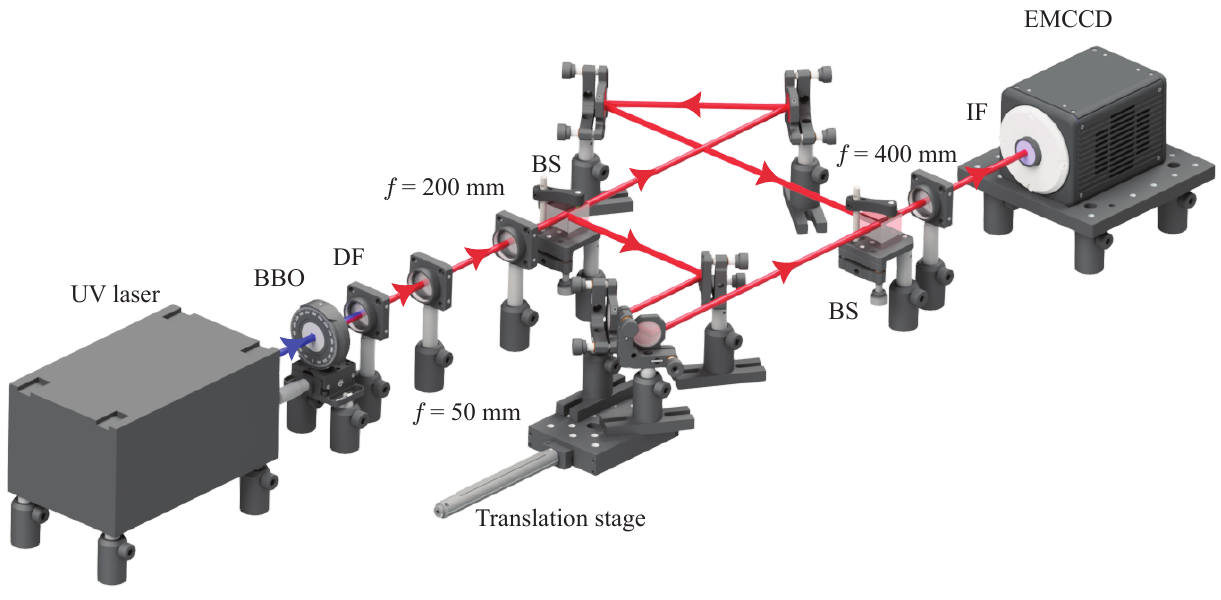}
    \caption{\textbf{Schematic of the experimental setup.} UV laser of central wavelength 355 nm; BBO, $\beta-$barium borate crystal; DF, dichroic filter; BS, beam splitter; IF, interference filter of central wavelength 710 nm with bandwidth of 10 nm.}
    \label{fig:Experimental_Setup}
\end{figure*}

\section{Experimental Observations}
We employ the interferometric technique demonstrated in Ref.~\cite{kulkarni2017single} to measure the Schmidt spectrum for different crystal thicknesses and phase-matching angles. 
The experimental setup used for measuring the angular Schmidt spectrum is shown in Fig.~\ref{fig:Experimental_Setup}. A UV pump laser with wavelength $\lambda_p = 355\,\mathrm{nm}$ and beam waist $w_p = 507\,\mu\mathrm{m}$ is used to generate type-I SPDC in a $\beta$-barium borate (BBO) crystal. The crystal is mounted on a goniometer with angular resolution $0.04^\circ$, allowing control of the phase-matching angle $\theta_p$. A dichroic mirror (DM) is used to block the UV pump beam while allowing the down-converted photon pairs to pass through. The down-converted photons are directed into a Mach-Zehnder-type interferometer, and for a given setting of crystal thickness and phase matching angle, interferograms are recorded at two phase settings differing by $\pi$, using an Andor iXon Ultra EMCCD camera (512 $\times$ 512 pixels) with an acquisition time of $2\,\mathrm{s}$.
 \begin{figure}[h!]
     \centering
     \includegraphics[
    width=\columnwidth,
    trim=7 8 7 7,
    clip
    ]{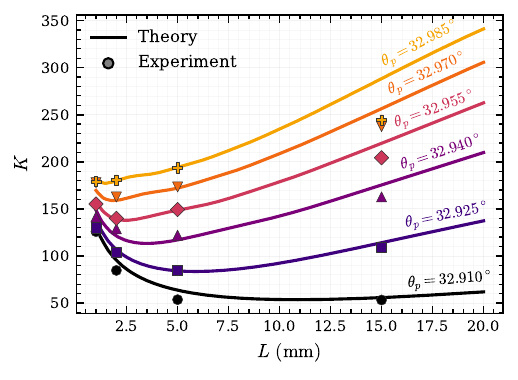}
     \caption{\textbf{Dependence of the Schmidt number on crystal thickness for various phase-matching angles.} Solid curves represent theoretical predictions based on the exact phase-matching formulation, while markers denote experimental measurements. The results show a non-monotonic dependence of the Schmidt number on crystal thickness.}
     \label{fig:KvsL}
 \end{figure}
From the measured angular Schmidt spectrum, we determine the Schmidt number \(K\). Figure~\ref{fig:KvsL} shows the variation of $K$ with crystal thickness for different phase-matching angles. The solid curves represent theoretical predictions based on the exact phase-matching formulation, while the markers correspond to experimental data. For each phase-matching angle, the Schmidt number exhibits a non-monotonic dependence on crystal thickness: it initially decreases up to a certain thickness and subsequently increases for larger crystal thicknesses. The experimental measurements closely follow the theoretical trends and reproduce the predicted non-monotonic dependence of the Schmidt number on crystal thickness, demonstrating the importance of spatial walk-off in determining the crystal-thickness dependence of the angular Schmidt spectrum. To the best of our knowledge, these measurements constitute the first experimental demonstration of a non-monotonic behavior of the Schmidt number on crystal thickness.

\section{Conclusion}
In conclusion, we have investigated the dependence of the angular Schmidt spectrum on crystal thickness in spontaneous parametric down-conversion. We have experimentally demonstrated that the Schmidt number exhibits a non-monotonic dependence on crystal thickness. We have shown that this behavior originates from spatial walk-off and is captured only when the complete phase-matching function is retained. In contrast, widely used approximate phase-matching models predict a monotonic decrease of the Schmidt number with crystal thickness and fail to reproduce the observed behavior. These results demonstrate the important role of spatial walk-off in determining the dimensionality of OAM-entangled states and highlight the need for accurate phase-matching models for reliable characterization of high-dimensional OAM entanglement. This can have important implications for high-dimensional state generation in high-flux applications where thick crystals are used.

\section{Acknowledgement}
We acknowledge financial support from the Science and Engineering Research Board through grants STR/2021/000035 and CRG/2022/003070, and from the Department of Science \& Technology, Government of India through the QuEST grant DST/ICPS/QuST/Theme-I/2019 and the National Quantum Mission (NQM) grant DST/FFT/NQM/QSM/2024/3 for quantum imaging.

\appendix
\renewcommand{\thefigure}{A\arabic{figure}}
\setcounter{figure}{0}

\section{Limitations of approximate phase-matching formulations}
\label{appendix:approximation}
The results presented in the main text demonstrate that retaining the complete phase-matching function is essential for correctly describing the angular Schmidt spectrum. Since several widely used theoretical formulations~\cite{torres2003quantum,miatto2011full,yao2011angular,miatto2012bounds,miatto2012cartesian} are based on approximate phase-matching conditions, it is instructive to examine how these approximations influence the predicted coincidence amplitudes and Schmidt spectrum. In this appendix, we compare the predictions obtained using the approximate phase-matching formulation of Ref.~\cite{miatto2011full} with those obtained when the exact phase-matching function is retained.

\subsection{Comparison of zero-radial-mode spectra}\label{appendix:zero_mode}
Reference~\cite{miatto2011full} derived expressions for the coincidence amplitudes in both the thick- and thin-crystal limits. Their derivation assumes collinear phase matching and identical refractive indices for the pump, signal, and idler fields inside the nonlinear crystal. Under these assumptions, the longitudinal phase mismatch reduces to the approximate expression given by Eq.~(\ref{eq:deltak_approx}). Using their numerical analysis, they concluded that the thick- and thin-crystal formulations remain in excellent agreement even for crystal thicknesses extending to several tens of centimeters.

The coincidence amplitude describes the probability amplitude for detecting the signal and idler photons in specific Laguerre--Gaussian modes characterized by radial indices \(p_s\) and \(p_i\) and azimuthal indices \(l_s=l\) and \(l_i=-l\). Since the angular Schmidt spectrum is obtained from these coincidence amplitudes after summing over all radial modes, their accurate evaluation is essential for correctly characterizing OAM entanglement. We therefore begin by comparing the probability distribution obtained from these coincidence amplitudes using the approximate and exact phase-matching models. Within the approximate phase-matching model, the coincidence amplitude in the thick-crystal regime is given by~\cite{miatto2011full}
\begin{align}\label{thickcrys}
C_{p_s,p_i}^{l,-l} &\propto K_{p_s,p_i}^{|l|}\int_{-L/2}^{L/2}dt \frac{(2B)^{|l|}(1 - \frac{4I}{T})^{p_s}(1 - \frac{4S}{T})^{p_i}}{T^{|l|+1}} \nonumber \\
&\quad\times 2F_1\left[\substack{-p_i,-p_s,\\-p_i-p_s-|l|};\frac{T(T-4I-4S+4)}{(T-4S)(T-4I)}\right],
\end{align}
where ${}_2F_1$ denotes the Gaussian hypergeometric function, and the remaining parameters are defined in Ref.~\cite{miatto2011full}. The dependence on crystal thickness enters explicitly through the integration over the crystal length. In the thin-crystal limit \((L\rightarrow0)\), the integral can be evaluated analytically, resulting
\begin{align}\label{thincrys}
C_{p_s,p_i}^{l,-l} &\propto K_{p_s,p_i}^{|l|} \frac{(1-\gamma_s^2+\gamma_i^2)^{p_s}(1-\gamma_i^2+\gamma_s^2)^{p_i}(-2\gamma_s\gamma_i)^{|l|}}{(1+\gamma_s^2+\gamma_i^2)^{p_s+p_i+|l|}} \nonumber \\
&\quad\times 2F_1\left[\substack{-p_i,-p_s,\\-p_i-p_s-|l|};\frac{1-(\gamma_s^2+\gamma_i^2)^2}{1-(\gamma_s^2-\gamma_i^2)^2}\right],
\end{align}
where \(\gamma_s\) and \(\gamma_i\) denote the ratios of the pump-beam waist to the signal and idler beam waists, respectively. In this limit, the explicit dependence on crystal thickness disappears and the coincidence amplitude depends only on the beam parameters. Equations~(\ref{thickcrys}) and~(\ref{thincrys}) constitute the principal results of Ref.~\cite{miatto2011full}. Based on these expressions, the authors concluded that the thin-crystal approximation remains valid over a wide range of crystal thicknesses, extending up to several tens of centimeters.

To assess the validity of the approximate phase-matching model, we compare the probability distribution corresponding to the zero-radial modes obtained using the approximate and exact phase-matching functions. We consider the zero-radial-mode probability distribution

\begin{align}
P_{0,0}^{l,-l}
=
\left|C_{0,0}^{l,-l}\right|^2,
\label{Eq:Spectrum_zero_mode}
\end{align}

which corresponds to the probability of detecting the signal and idler photons in the Laguerre--Gaussian modes with radial indices \(p_s=p_i=0\) and OAM indices \(l\) and \(-l\), respectively. The zero-radial-mode spectrum is of particular experimental relevance since many OAM measurement techniques predominantly detect the lowest radial mode.
\begin{figure}[h!]
     \centering
     \includegraphics[width=\columnwidth,trim=7 8 7 7,clip]{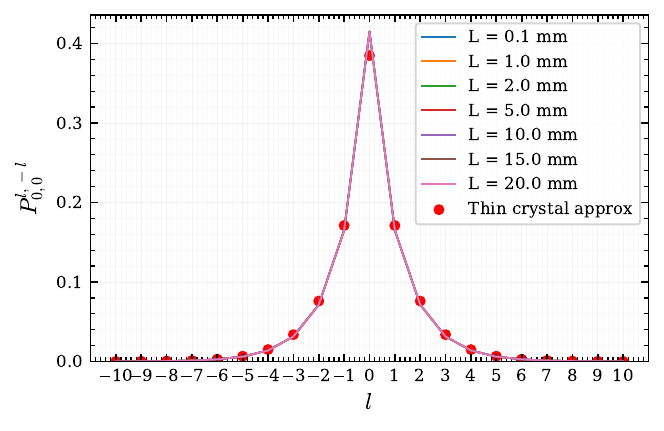}
     \caption{\textbf{Comparison of the zero-radial-mode spectrum obtained using the approximate phase-matching model.} The probability distribution \(P_{0,0}^{l,-l}\) is calculated using the thick-crystal expression [Eq.~(\ref{thickcrys})] and compared with the thin-crystal approximation [Eq.~(\ref{thincrys})] for different crystal thicknesses. The two formulations remain in excellent agreement over a wide range of crystal thicknesses.}
     \label{fig:approx}
 \end{figure}
 
Using the coincidence amplitudes given by Eqs.~(\ref{thickcrys}) and (\ref{thincrys}), we compute the zero-radial-mode spectrum for different crystal thicknesses. The results are shown in Fig.~\ref{fig:approx}. The spectrum obtained from the thin-crystal expression [Eq.~(\ref{thincrys})] remains in excellent agreement with that calculated using the thick-crystal expression [Eq.~(\ref{thickcrys})] over the entire range of crystal thicknesses considered. These numerical results are consistent with the conclusions of Ref.~\cite{miatto2011full}, which reported that the thin-crystal approximation remains valid even for crystal lengths extending to several tens of centimeters.
\begin{figure}[h!]
     \centering
     \includegraphics[width=\columnwidth,trim=7 8 7 7,clip]{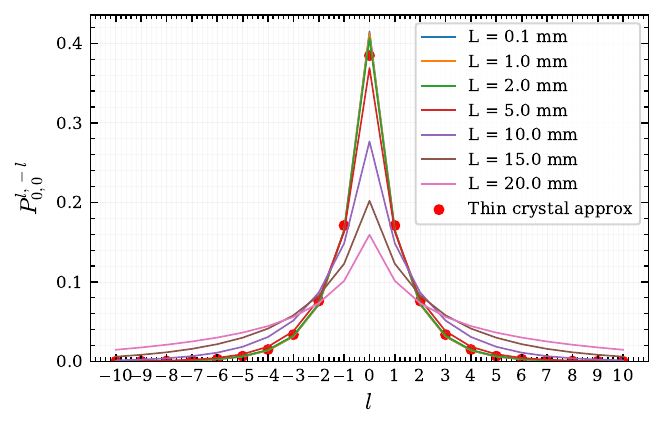}
     \caption{\textbf{Comparison of the zero-radial-mode spectrum obtained using the exact phase-matching function.} The probability distribution broadens significantly with increasing crystal thickness, demonstrating that the thin-crystal approximation no longer reproduces the behavior predicted by the complete phase-matching model.}
     \label{fig:unapprox}
 \end{figure}
 
To examine whether this behavior is a consequence of the approximate phase-matching model, we repeat the same calculation using the exact phase-matching function derived from Eqs.~(\ref{eq:phasematching}) and (\ref{eq:k's}), while retaining all other experimental parameters unchanged. The corresponding spectra are shown in Fig.~\ref{fig:unapprox}. In contrast to Fig.~\ref{fig:approx}, the zero-radial-mode spectrum broadens noticeably with increasing crystal thickness. Consequently, the thin-crystal approximation no longer reproduces the spectrum predicted by the exact phase-matching model.

This comparison demonstrates that the apparent agreement between the thick- and thin-crystal formulations reported in Ref.~\cite{miatto2011full} arises from the approximate phase-matching model employed in their derivation. When the complete phase-matching function is retained, the crystal-thickness dependence changes qualitatively, indicating that the thin-crystal approximation is no longer adequate for describing thicker crystals, where spatial walk-off becomes significant.

\subsection{Comparison of angular Schmidt spectra}\label{appendix:schmidt_comparison}
While the comparison presented above is useful for assessing the zero-radial-mode spectrum, the quantity of primary interest in the present work is the angular Schmidt spectrum, which is obtained by summing over all radial modes. Within the approximate formulation of Ref.~\cite{miatto2011full}, the angular Schmidt spectrum is evaluated as

\begin{align}
S_l
=
\sum_{p_s=0}^{\infty}
\sum_{p_i=0}^{\infty}
\left|C_{p_s,p_i}^{l,-l}\right|^2,
\label{eq:Schmidt_approx}
\end{align}

where the coincidence amplitudes are given by Eq.~(\ref{thincrys}). In practice, the infinite summations over the radial indices must be truncated to a finite number of modes for numerical evaluation. Consequently, the calculated Schmidt spectrum may depend on the chosen truncation, making it important to examine the convergence of the approximate formulation.
\begin{figure}[h!]
     \centering
     \includegraphics[width=\columnwidth,trim=7 11 8 7,clip]{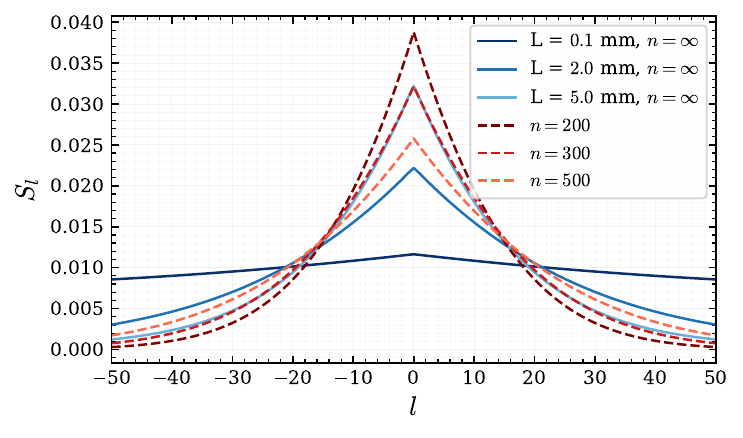}
     \caption{\textbf{Comparison of angular Schmidt spectra obtained using the approximate and exact formulations.}
    The spectra obtained from the approximate formulation [Eq.~(\ref{eq:Schmidt_approx})] are shown for different radial-mode truncation limits. As progressively higher radial modes are included, the spectrum becomes increasingly flat, indicating the lack of convergence of the approximate formulation. The spectrum obtained using the exact formulation [Eq.~(\ref{eq:schmidt_spectrum})], which analytically accounts for all radial modes, remains well behaved.}
     \label{fig:Schmidt plot}
 \end{figure}
 
To investigate this issue, we compute the Schmidt spectrum by retaining different numbers of radial modes. Specifically, we consider truncation limits of \(n=200\), \(300\), and \(500\) for both the signal and idler fields. The resulting spectra are shown in Fig.~\ref{fig:Schmidt plot}. Rather than approaching a converged solution, the spectrum becomes progressively flatter as the number of included radial modes increases, indicating that increasingly higher-order OAM modes acquire nearly equal probabilities. This behavior indicates that the Schmidt spectrum obtained from the approximate formulation does not converge with increasing radial-mode truncation.

For comparison, Fig.~\ref{fig:Schmidt plot} also shows the angular Schmidt spectrum obtained using the formulation of Ref.~\cite{kulkarni2018angular}, given by Eq.~(\ref{eq:schmidt_spectrum}), which retains the complete phase-matching function and analytically accounts for all radial-mode contributions. In contrast to the approximate formulation, the resulting Schmidt spectrum remains well behaved and exhibits a physically meaningful dependence on crystal thickness.

The comparison highlights a fundamental limitation of the approximate phase-matching formulation. Although it provides compact analytical expressions for the coincidence amplitudes, the corresponding Schmidt spectrum suffers from convergence difficulties arising from the explicit summation over infinitely many radial modes. By analytically incorporating the radial-mode contributions while retaining the complete phase-matching function, the formulation of Ref.~\cite{kulkarni2018angular} overcomes these limitations and provides a reliable description of the angular Schmidt spectrum for arbitrary crystal thicknesses. 

The results presented in this appendix provide additional evidence that retaining the complete phase-matching function is essential for accurately describing both the coincidence amplitudes and the angular Schmidt spectrum of OAM-entangled photon pairs.

\bibliographystyle{apsrev4-2}
\bibliography{references}

\end{document}